\documentclass[namedreferences]{solarphysics}
\usepackage[optionalrh,solaromanenum]{spr-sola-addons} 
\usepackage{graphicx}                    
\usepackage{color}                       
\usepackage{url}                         

\begin{document}

\begin{article}

\begin{opening}

\title{The 22-Year Hale Cycle in Cosmic Ray Flux - Evidence for Direct Heliospheric Modulation}

%
\author{S. R.~\surname{Thomas}$^{1}$\sep
        M. J.~\surname{Owens}$^{1}$\sep
        M.~\surname{Lockwood}$^{1}$      
       }

%

%
  \institute{$^{1}$ University of Reading
                     email: \url{s.r.thomas@pgr.reading.ac.uk}
             }

\begin{abstract}

The ability to predict times of greater galactic cosmic ray (GCR) fluxes is important for reducing the hazards caused by these particles to satellite communications, aviation, or astronauts. The 11-year solar cycle variation in cosmic rays is highly correlated with the strength of the heliospheric magnetic field. Differences in GCR flux during alternate solar cycles yield a 22-year cycle, known as the Hale Cycle, which is thought to be due to different particle drift patterns when the northern solar pole has predominantly positive (denoted a qA$>$0 cycle) or negative (qA$<$0) polarities. This results in the onset of the peak cosmic ray flux at Earth occurring earlier during qA$>$0 cycles than for qA$<$0 cycles and hence the peak being more domed for qA$>$0 and more sharply peaked for qA$<$0. In this study, we demonstrate that properties of the large-scale heliospheric magnetic field are different during the declining phase of the qA$<$0 and qA$>$0 solar cycles, when the difference in GCR flux is most apparent. This suggests that particle drifts may not be the sole mechanism responsible for the Hale Cycle in GCR flux at Earth. However, it is also demonstrated that these polarity-dependent heliospheric differences are evident during the space-age but much less clear in earlier data: using geomagnetic reconstructions, it is shown that for the period of 1905 - 1965, alternate polarities do not give as significant a difference during the declining phase of the solar cycle. Thus we suggest that the 22-year cycle in cosmic ray flux is at least partly the result of direct modulation by the heliospheric magnetic field and that this effect may be primarily limited to the grand solar maximum of the space-age.

\end{abstract}

%

\end{opening}

%

The final publication is available at http://link.springer.com using the following link: http://link.springer.com/article/10.1007/s11207-013-0341-5

\section{Introduction}

During the recent solar minimum, which was longer and deeper than others observed for over a century \cite{Lockwood10}, galactic cosmic ray (GCR) flux has reached its highest values of the space-age \cite{Mewaldt10}. The high GCR flux has implications for satellites, space craft and aviation ({\it e.g.}, \inlinecite{Hapgood10}) due to their high energies, making the ability to predict times of greater fluxes of cosmic rays critical for mission planning and reducing such hazards. It is also important to study the propagation and modulation of GCRs throughout the heliosphere for the purposes of long-term reconstructions of solar parameters ({\it e.g.} \opencite{McCracken04}; \opencite{Usoskin11}). Indeed, cosmogenic isotopes generated in the ￼￼atmosphere by GCRs and stored in dateable terrestrial reservoirs such as ice sheets and tree trunks, are our only source of information on solar variability on millennial timescales \cite{Beer06}. For the present study, GCR flux is inferred using high-latitude ground-based neutron monitors. As GCRs enter the terrestrial atmosphere, they collide with atmospheric particles, producing secondary particles such as neutrons, which are then observed at the detectors situated around the globe. The neutron monitor used for this study is at McMurdo, Antarctica which is run by the Bartol Institute and has been recording data since 1964. The cut-off rigidity for this neutron monitor, set by the geomagnetic field, is smaller than that set by the atmosphere, due to the strength and more-vertical orientation in polar regions of the geomagnetic field. This means that the instrument responds to energies down to about 1 GeV, whereas a station near the Equator (where the cut-off rigidity is set by the geomagnetic field) will respond to particles of energy exceeding about 16 GeV. The fractional modulation of cosmic rays by the heliosphere is greater at lower energies and hence by selecting this high-latitude station we detect the larger variation of the lower energy particles \cite{Bieber04}.

\inlinecite{Schwabe1843} was the first to recognise the 11-year solar cycle variation using the periodicity in sunspot number records and the signature of this variation in cosmic rays was detected using ionisation chambers by \inlinecite{Forbush54}. Evidence for the 22-year solar cycle variation was first reported on by \inlinecite{Ellis1899}, who observed high counts of geomagnetically quiet days during alternate minima in the 1850s and 1870s. \inlinecite{Chernosky66} showed further characteristic differences in geomagnetic activity in alternate 11-year cycles. The "odd" and "even" numbered solar cycles have been shown to be different in cosmic ray fluxes at Earth (e.g., \opencite{Webber88}) giving a 22-year cycle, known as the Hale Cycle, as also seen in sunspot polarity and latitude (\opencite{Hale25}). \inlinecite{Allen00} compared neutron counts with sunspot numbers for solar activity cycles 19-22 inclusive. He produced "modulation cycles" where each year a data point is plotted, between sunspot number and neutron counts, which map out an approximately circular pattern throughout the cycle. They showed that the shape of these plots were vastly different between the odd solar cycles 21 and 23 and the even cycles 20 and 22. He also noted that as sunspot numbers increase after solar minimum, the cosmic ray flux drops quicker for "odd" cycles than for "even" cycles.

Studies of the 22-year cycle in cosmic ray fluxes from modern neutron monitors (e.g., \opencite{Webber88}; \opencite{Smith90}) have lead to the description of neutron counts following an alternate "flat-topped" and "peaked" pattern. The polarity of the solar field [A] is taken to be negative when the dominant polar field is inward in the northern hemisphere and outward in the southern (e.g. \opencite{Ahluwalia10}; and references therein) and positive if the opposite is true. Curvature and gradient drift directions are reversed if the sign of the charge of the particle [q] is reversed, so it is customary to define cycles by the polarity of the product [qA]. The occurrence of "flat-topped" and "peaked" maxima has been found to be in agreement with the expected effect of curvature and gradient drifts of cosmic ray protons (\opencite{Jokipii77}; \opencite{Jokipii81}; \opencite{Potgieter95}; \opencite{Ferreira04}): during cycles with positive polarity (qA$>$0), cosmic ray protons arrive at Earth after approaching the poles of the Sun in the inner heliosphere and moving out along the heliospheric current sheet (HCS). Conversely, during negative polarities (qA$<$0), cosmic ray protons approach the Sun along the HCS plane and leave via the poles. The ease with which cosmic rays can travel toward Earth along the HCS during qA$<$0 cycles is thought to depend on the HCS tilt (or inclination) relative to the solar Equator. Shielding of GCRs is also provided by scattering of particles off irregularities in the heliospheric field. Because the number and size of these irregularities tends to scale with the field strength, this is well quantified by the open solar flux (OSF) (\opencite{Rouillard04}), the total magnetic flux leaving the coronal source surface (usually defined to be at a heliocentric distance of 2.5 R$_{\odot}$, where R$_{\odot}$ is a mean solar radius). Surveys of {\it in-situ} data show that the near-Earth interplanetary medium also displays 22-year cycles ({\it e.g.} \opencite{Hapgood91}) and the results of \inlinecite{Rouillard04} suggest that the 22-year variation was primarily caused by that in heliospheric field strength, with less influence of drift effects than previously thought.

The polar field reversal, which must separate qA$>$0 and qA$<$0 cycles, occurs at, or just after, sunspot maximum for each solar cycle. It is triggered by magnetic flux migrating up from sunspot groups towards the poles which cancels out the pre-existing flux of opposite polarity already situated here \cite{Harvey96}. This behaviour is clearly visible in photospheric magnetogram data.

The tilt angle of the heliospheric current sheet (HCS) has been shown to be a key parameter in the modulation of cosmic rays. A model proposed by \inlinecite{Alanko07} suggested that the modulation can be described by a combination of the HCS tilt angle, the Sun’s polarity and the unsigned open solar flux (OSF). This model gives good agreement throughout solar cycles 19-23. \inlinecite{Cliver01} compared the heliospheric tilt angle for solar cycle 21, 22, and the available data of solar cycle 23 at the time. They noted that during the declining phase of solar cycle there was a more gradual decay of the HCS tilt angle following the odd cycle than the even cycle. During the ascending phase however, both cycles were remarkably similar. From this they concluded that this is likely to be due to differences in the evolution of the large-scale magnetic field on the decay of the solar cycle. This study was updated by \inlinecite{Cliver11} who added the HCS tilt angle data for the remainder of solar cycle 23. They found that the recent cycle was indeed similar to cycle 21 in shape, but that solar cycle 23 was much longer. In this paper, we build on the studies of \inlinecite{Cliver01} and \inlinecite{Cliver11} by evaluating other heliospheric parameters to investigate the possible difference in the declining phase of the solar cycle, in particular whether the difference between GCR flux in qA$<$0 and qA$>$0 cycles has its origin in differences in the heliospheric field strength, not just in its direction (as would be expected for drift effects alone).


\section{The 22-Year Solar Cycle Variations}

In this study, we consider “polarity cycles” as the intervals between polar polarity reversals ({\it i.e.} solar maximum to solar maximum) rather than the conventional solar cycle ({\it i.e.} from solar minimum to solar minimum), as has previously been studied. This enables us to better isolate effects of solar polarity. Thus, we assign a phase [$\epsilon$$_p$], varying linearly between 0$^\circ$ and 360$^\circ$ between the two polarity reversals (such that its relationship to the sunspot cycle phase [$\epsilon$] defined from solar minimum to minimum by \inlinecite{Lockwood12} is $\epsilon$$_p$ $\approx$ $\epsilon$ - 2$\pi$x(4.5$/$L), where L is the solar cycle length in years). However, polarity reversals are not simple to define from photospheric magnetogram data, as there are annual fluctuations in observed polar polarities due to the inclination of the ecliptic plane with respect to the heliographic equator \cite{Babcock55}. Furthermore, polarity reversals do not occur simultaneously at both poles but are often separated by over a year \cite{Babcock59}. The present study used polarity reversal times for solar cycles 21-23 from \inlinecite{Svalgaard05} and \inlinecite{Hathaway12}, an extension to the polarity reversal of solar cycle 24 (see \opencite{Lockwood12}) and analysis from \inlinecite{Babcock59} for solar cycle 20.

Within one solar hemisphere, the mean of the earliest and latest times at which the polarity reversal may have occurred are defined as the times at which the average field in that polar region crosses zero. These are displayed on Figure 1 as the vertical red lines. The top panel shows these data in years since solar minimum, defined as the time of rapid increase in the average sunspot latitude \cite{Owens11a}. The bottom panel shows the same polarity reversal data as a function of solar cycle phase [$\epsilon$], defined as 0$^\circ$ at the start of the solar cycle and 360$^\circ$ at the end of the solar cycle, which effectively normalises for the variable length of solar cycles. The black crosses in figure 1 are times when the average north minus average south polar fields cross zero (this data however is unavailable prior to cycle 20 and so these crosses are not included in Figure 1) This is generally used as a measure of the global solar dipole having reversed ({\it e.g.} \opencite{Hathaway12}).

\begin{figure}[h] 
\centerline{\includegraphics[width=0.7\textwidth,clip=]{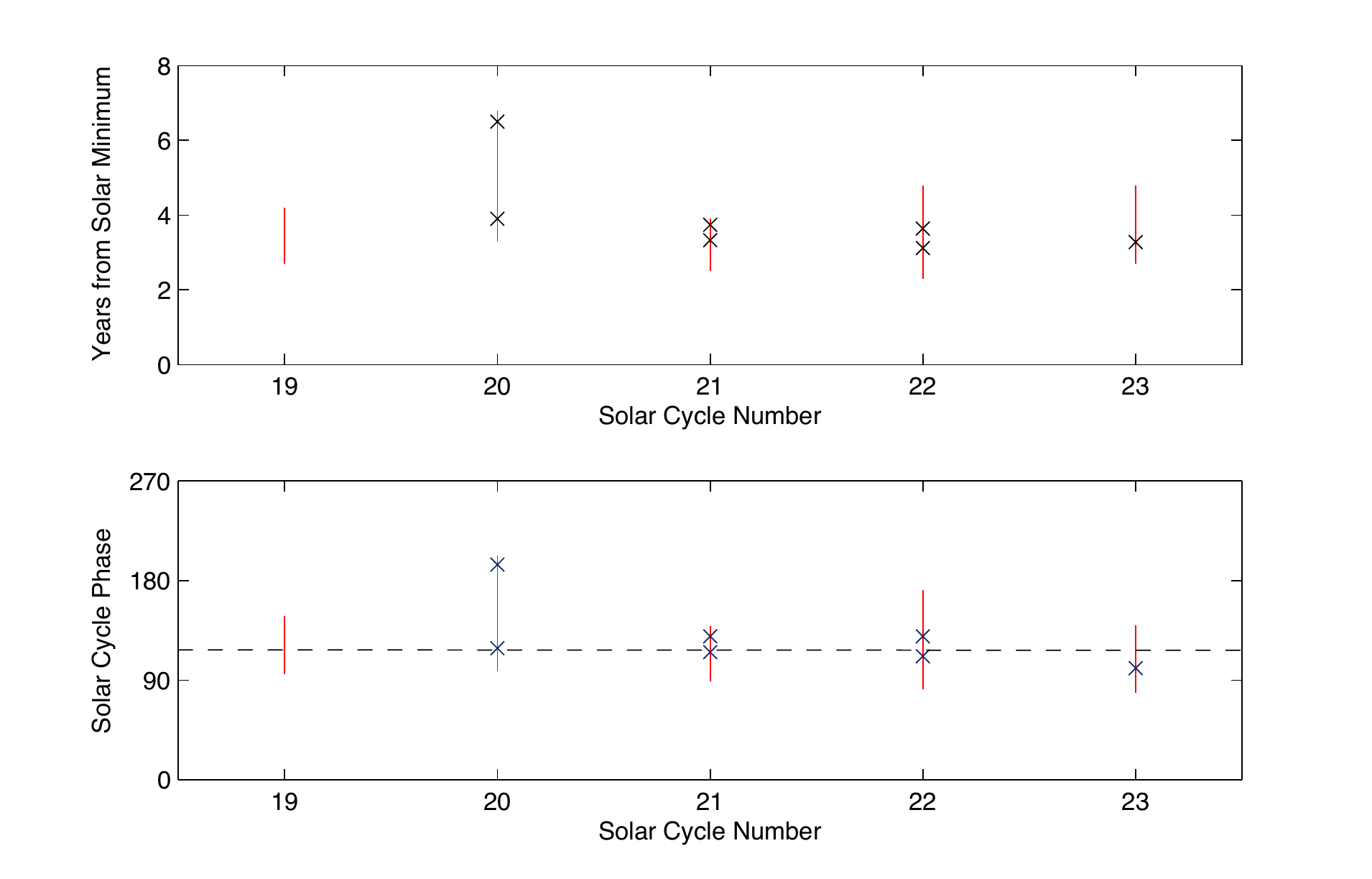}}
\caption{Solar polarity reversal times for solar cycles 19 to 23, as estimated from photospheric magnetograms. The red lines show the earliest and latest times that the average field of either the north or the south magnetic polar region crosses zero. Black crosses give best estimates for polarity reversals. The top panel shows the times from sunspot minimum, in years, while the bottom panel shows solar cycle phase [$\epsilon$]. The dashed horizontal line in the bottom panel show the value used here to determine the timing of polarity reversals from sunspot data.}
\label{fig:1}
\end{figure}

In order to apply this analysis to pre-space-age solar cycles, it is necessary to be able to estimate the time of polarity reversals without the aid of photospheric magnetograms, instead relying only on sunspot data. We do this by calculating the solar cycle phase, [$\epsilon$], which is the best fit through all the potential times of polarity reversal shown on Figure 1. We find a phase of $\epsilon$ $=$ 125$^\circ$, as shown by the horizontal dashed line and this $\epsilon$ is used to define $\epsilon$$_p$ $=$ 0$^\circ$. The error on the phase is 20$^\circ$ or an average of 0.5 to 1 year in the top panel. The blue (qA$>$0) and red (qA$<$0) lines on Figure 2 are based on polarity reversal timings approximated by this method. The grey shaded regions are the full extents of the polarity reversal times estimated from photospheric magnetograms. In general, there is good agreement between the two methods. This is because, while solar cycle length can vary considerably, it tends to be a result of short or long declining phases, with rise phases showing much less variability in length (\opencite{Waldmeier35}; \opencite{Hathaway94}; \opencite{Owens11a}).

\begin{figure}[h]\centerline{\includegraphics[width=0.8\textwidth,clip=]{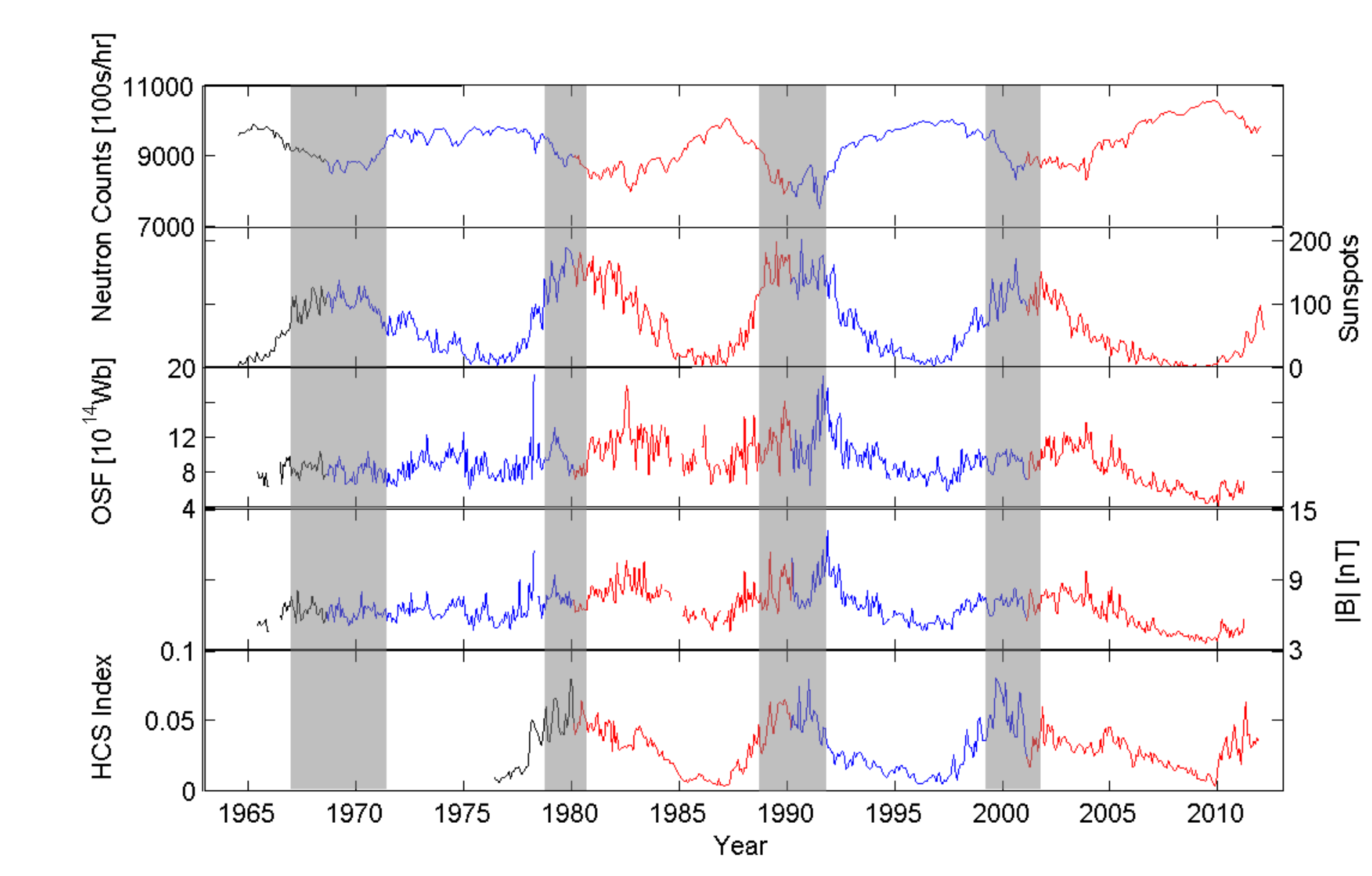}}
\caption{Time series (from top):- neutron monitor counts at McMurdo; sunspot number; unsigned open solar flux; magnitude of the heliospheric magnetic field in near-Earth space; and heliospheric current sheet tilt index. The 22-year solar cycle is clearly seen in the cosmic ray count rates. The times of qA$<$0 polarity are shown in red and qA$>$0 in blue with the change in colours representing the polarity reversal time estimated using sunspot data. The grey boxes represent the times of polarity reversals estimated from photospheric magnetograms as described in the text.}
\label{fig:2}
\end{figure}

The top panel of Figure 2 shows neutron monitor counts at McMurdo. The 22-year cycle in cosmic ray flux, with the alternate "flat-topped" (blue) and "peaked" (red) pattern, is clearly visible. The second panel shows sunspot number. The third panel shows the unsigned open solar flux (OSF), calculated from 4$\pi$AU$^2$$|$B$_R$$|$, where AU is the Earth-Sun distance and B$_R$ is the daily mean radial magnetic field from the OMNI dataset \cite{King05}. The bottom panel shows the HCS index, a useful parameter for quantifying any tilt and the warped nature of the heliospheric current sheet (in other studies often quantified by “tilt angle”). It is found by applying a uniform grid across the magnetogram-constrained potential field source surface (PFSS) and computing the fraction of grid boxes that have the opposite polarity to their immediate longitudinal neighbour \cite{Owens11b}. At solar maximum, much of the HCS is highly inclined with the rotation axis and it is highly warped due to a strong quadrupole moment, giving a large HCS index value. At solar minimum, when the quadrupole moment is weaker and the dipole more rotationally aligned, the HCS index has a much lower value. As discussed above, the HCS has been found to play a key role in the modulation of cosmic rays ({\it e.g.} \opencite{Smith86}) as distortions in the HCS are associated with corotating interaction regions (CIRs), which can act as shields to GCR propagation (\opencite{Rouillard07}). Structures in the heliospheric magnetic field can result in GCRs experiencing drift effects or scattering off of irregularities ({\it e.g.} \opencite{Parker65}; \opencite{Jokipii77}). OSF is also included in a model by \inlinecite{Alanko07}, who found it to be strongly anti-correlated with neutron count rates ({\it e.g.} \opencite{Lockwood03}; \opencite{Rouillard04}). Note that the HCS inclination index data shows a more gradual decline during the declining phase of qA$<$0 cycles, compared to qA$>$0 cycles, as noted by \inlinecite{Cliver01} and \inlinecite{Cliver11}.


\section{Differences Between qA$<$0 and qA$>$0 Cycles}

To examine differences between heliospheric and GCR properties in qA$<$0 and qA$>$0 polarity cycles, we here use a superposed epoch (composite) analysis. As a demonstration of the analysis process, we first apply this analysis to neutron monitor count rates in Figure 3.

\begin{figure}[h]\centerline{\includegraphics[width=0.7\textwidth,clip=]{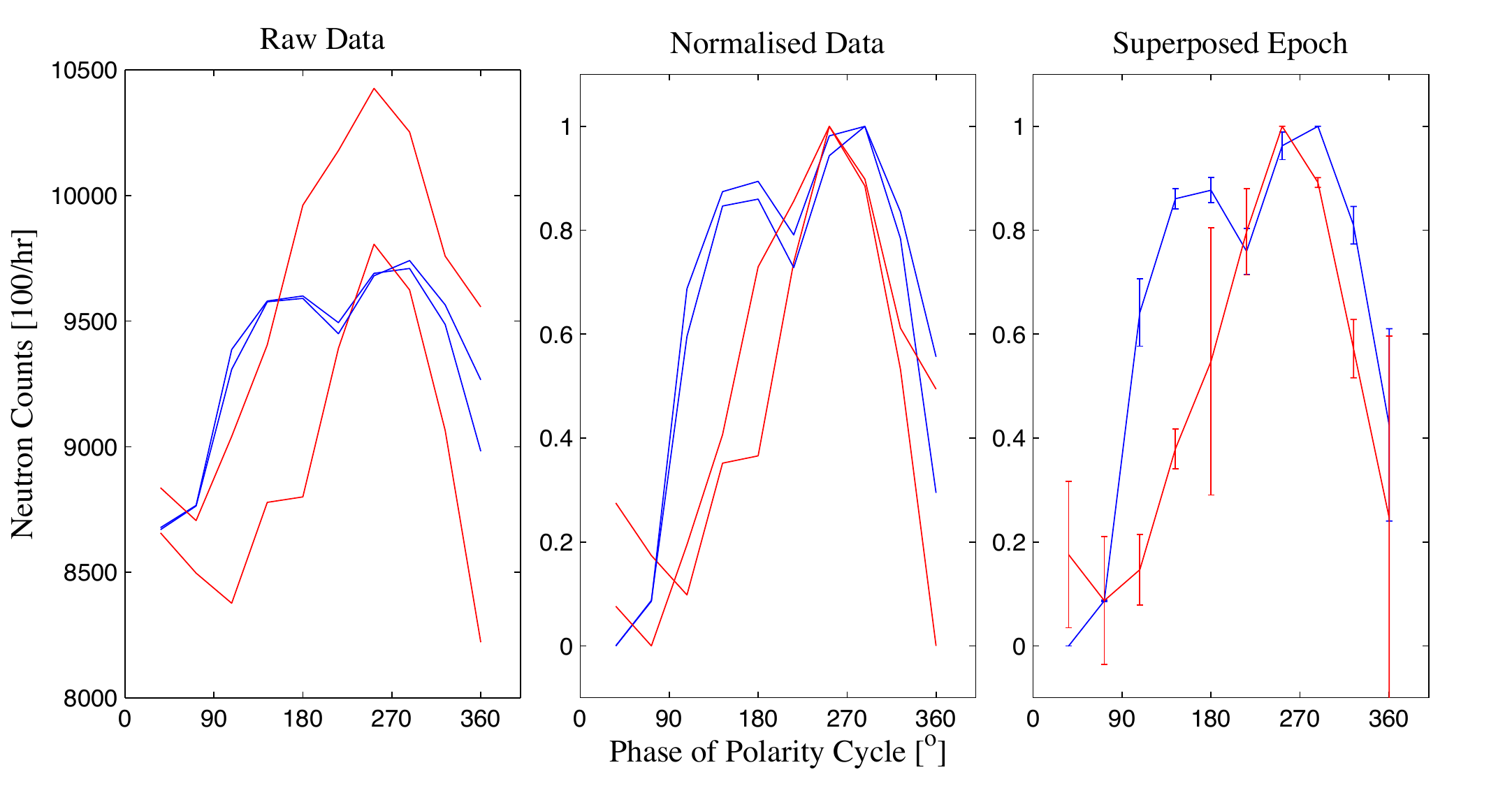}}
\caption{Neutron monitor counts as a function of polarity cycle phase [$\epsilon$$_p$], for qA$>$0 (blue) and qA$<$0 (red) cycles. From left: raw data, normalised data and superposed epoch (composite) analysis. The error bars are plus or minus one standard deviation.}
\label{fig:3}
\end{figure}

The red lines in Figure 3 represent qA$<$0 cycles while the blue lines are qA$>$0 cycles. The left panel shows the raw data (27 day means) of McMurdo neutron monitor count rates, as a function of the polarity phase, [$\epsilon$$_p$] (defined using the sunspot method of defining the polarity cycle start/end times). In the case of the anticipitated polarity reversal of cycle 24, we have used the date of 2.4 months into 2013, corresponding to a phase of 126$^\circ$ through the polarity cycle \cite{Lockwood12}. The middle panel shows the data normalised to the maximum and minimum values over that individual polarity cycle, in order to remove systematic cycle- to-cycle amplitude variations. Finally, the right panel shows the average for normalised parameters over qA$>$0 and qA$<$0 cycles, with error bars showing plus/minus one standard deviation.

The neutron monitor count rates clearly show the Hale cycle. The qA$<$0 and qA$>$0 cycles display differing shapes, with the “peaked” and “flat-top” profiles largely the result of differences in the first half of the polarity cycle ({\it i.e.}, the declining phase of the sunspot cycle), although there is a shorter, less pronounced difference after the cosmic ray peak ({\it i.e.}, the rising phase of the sunspot cycle). In Figure 4 we now repeat this analysis for a number of other solar and heliospheric parameters.

\begin{figure}[h]\centerline{\includegraphics[width=0.8\textwidth,clip=]{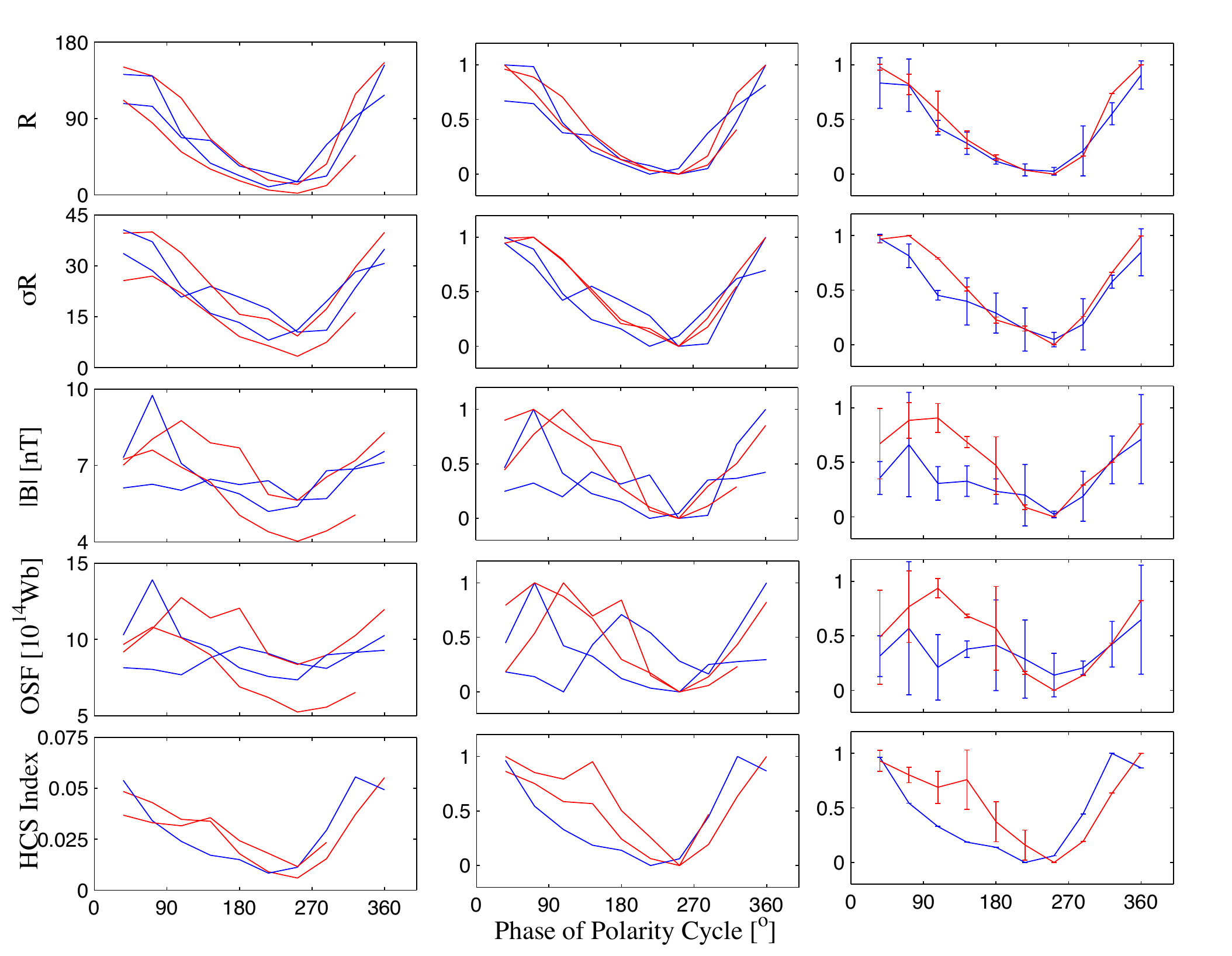}}
\caption{Panels from left to right give raw data, normalised data and averages of the qA$<$0 and qA$>$0 polarity cycles (in red and blue respectively). These plots are for (from top to bottom): monthly sunspot number [R], monthly standard deviation of daily sunspot number [$\sigma$$_R$], near-Earth magnetic field strength [$|$B$|$], open solar flux [OSF], and HCS inclination index. The error bars are plus and minus one standard deviation of the two polarity cycles used. Note that there is only data on the HCS inclination index for one qA$>$0 cycle and so no standard deviations can be given in the bottom right panel. The data has been averaged over bins in cycle phase [$\epsilon$$_p$], that are 36$^\circ$ wide.}
\label{fig:4}
\end{figure}

The format of Figure 4 is the same as for Figure 3, with raw data in the left column, normalised data in the middle and a superposed epoch analysis shown on the right. The four rows show (from top to bottom) the international sunspot number [R], the monthly standard deviation of daily sunspot number [$\sigma$$_R$], the near-Earth magnetic field strength [$|$B$|$], open solar flux [OSF] and the HCS inclination index. For the HCS inclination index, there is one cycle of data available for qA$>$0, so no error bars can be given.

The sunspot number shows little difference between qA$<$0 and qA$>$0 cycles, suggesting that the start/end times for the cycles have been well defined and there is no Hale effect in [R]. However, the standard deviation of the sunspot number does show evidence of a significant difference between the two polarities: in the second panel, we see a greater variability in sunspot number during qA$<$0 polarity cycles than during qA$>$0 cycles. The increased variability could be the result of “active longitudes” (e.g. \opencite{Ruzmaikin00}; \opencite{Berdyugina03}) separated by quiet longitudes. This is in agreement with an anticorrelation between cosmic ray flux and non-asymmetric open solar flux \cite{Wang06}, which is responsible for longitudinal structure in the heliosphere. It is worth noting that \inlinecite{Gil08} found that the 27-day variability of neutron monitor counts was greater in qA$>$0 cycles than in qA$<$0 cycles. However, this is the opposite behaviour to that for sunspot numbers noted here. This though could be a result of rotating compression regions associated with the HCS known as corotating interaction regions (CIRs), being more effective modulators during qA$>$0 cycles \cite{Richardson99}, rather than a change in CIR and/or heliospheric properties themselves.

A similar signature is also seen in the other three rows, which show the heliospheric magnetic field, OSF and the HCS inclination index. There is a significant difference between qA$<$0 and qA$>$0 cycles around $\epsilon$$_p$ = 125$^\circ$ (which corresponds to the declining phase of the solar cycle). The HCS inclination index result is also consistent with a greater prevalence of active longitudes during the declining phase of the solar cycle under qA$<$0 conditions. A very similar pattern is also noted for HCS tilt angle (not presented here). The only available qA$>$0 cycle is consistently outside of the error bars throughout the first half of the polarity cycle, the time when the qA$<$0 and qA$>$0 cosmic ray values differ most significantly. Hence, the result for cycle 23 is consistent with the previous findings of \inlinecite{Cliver01}.

Solar cycle 20 was unusual in terms of the magnitude of the near-Earth magnetic field and OSF being particularly “flat” and showing little solar-cycle variation. As can be seen from Figure 4, this cycle does indeed have an effect on the difference between the average behaviour of $|$B$|$ and OSF within the qA$>$0 and qA$<$0 cycles. However, we note that removing this cycle does not remove the significance in the difference between average qA$>$0 and qA$<$0 cycles.

We have also performed tests of sensitivity to changing the exact start and end times of the polarity cycles. Varying the boundaries between the times of the north and south polar reversals by an interval of 0.5 - 1 year (from the error on the phase in Figure 1), during the space age, does vary the average curves to some degree. However it does not remove the differences between the qA$>$0 and qA$<$0 cycles during the first half of the polarity cycle, which remains significant. This test is applied again and discussed further in the next section.


\section{Geomagnetic Reconstructions of the Pre-Space Age Heliosphere}

We now consider data from before the space age. Magnetic field magnitude and OSF can be reliably reconstructed back to at least 1905 using geomagnetic data ({\it e.g.} \opencite{Lockwood09}; \opencite{Lockwood11}). We use these data sets to examine the behaviour of the heliospheric magnetic field over six additional, pre-space age polarity cycles. This enables us to test whether this difference in heliospheric parameters during qA$>$0 and qA$<$0 polarity cycles is limited to the space-age, which spans the recent grand solar maximum (\opencite{Solanki04}; \opencite{Lockwood09}; \opencite{Lockwood12}), or whether it is a more persistent feature. Geomagnetic reconstruction of the heliospheric field is limited to yearly values because annual variations in factors such as the ionospheric conductivity and Earth’s dipole tilt influence the coupling between the solar wind and the geomagnetic field. Thus bin sizes are take to be approximately one year (precisely one year is not possible as we consider solar cycle phase, not time). Between the geomagnetic reconstructions and the OMNI data, heliospheric magnetic field magnitude and open solar flux have been shown to be consistent \cite{Lockwood11}. We therefore take it that these parameters agree during the space-age that geomagnetic reconstructions can be taken as representative of the heliospheric magnetic field throughout the period of 1905 - 2012.

Figures 5 and 6 give a similar analysis to Figure 4 in the same format as Figures 3 and 4, namely for the raw data (left column), normalised data (middle column) and means and standard deviations (right column) of the qA$<$0 and qA$>$0 polarity cycles (in blue and red, respectively) determined using the sunspot method of defining $\epsilon$$_p$. This analysis is given for (from top) $|$B$^*$$|$, the heliospheric magnetic field magnitude, and OSF$^*$, the open solar flux (where the asterisks denotes that values are reconstructed from geomagnetic data), along with monthly sunspot numbers and monthly standard deviation of the daily sunspot number.

\begin{figure}[h]\centerline{\includegraphics[width=0.8\textwidth,clip=]{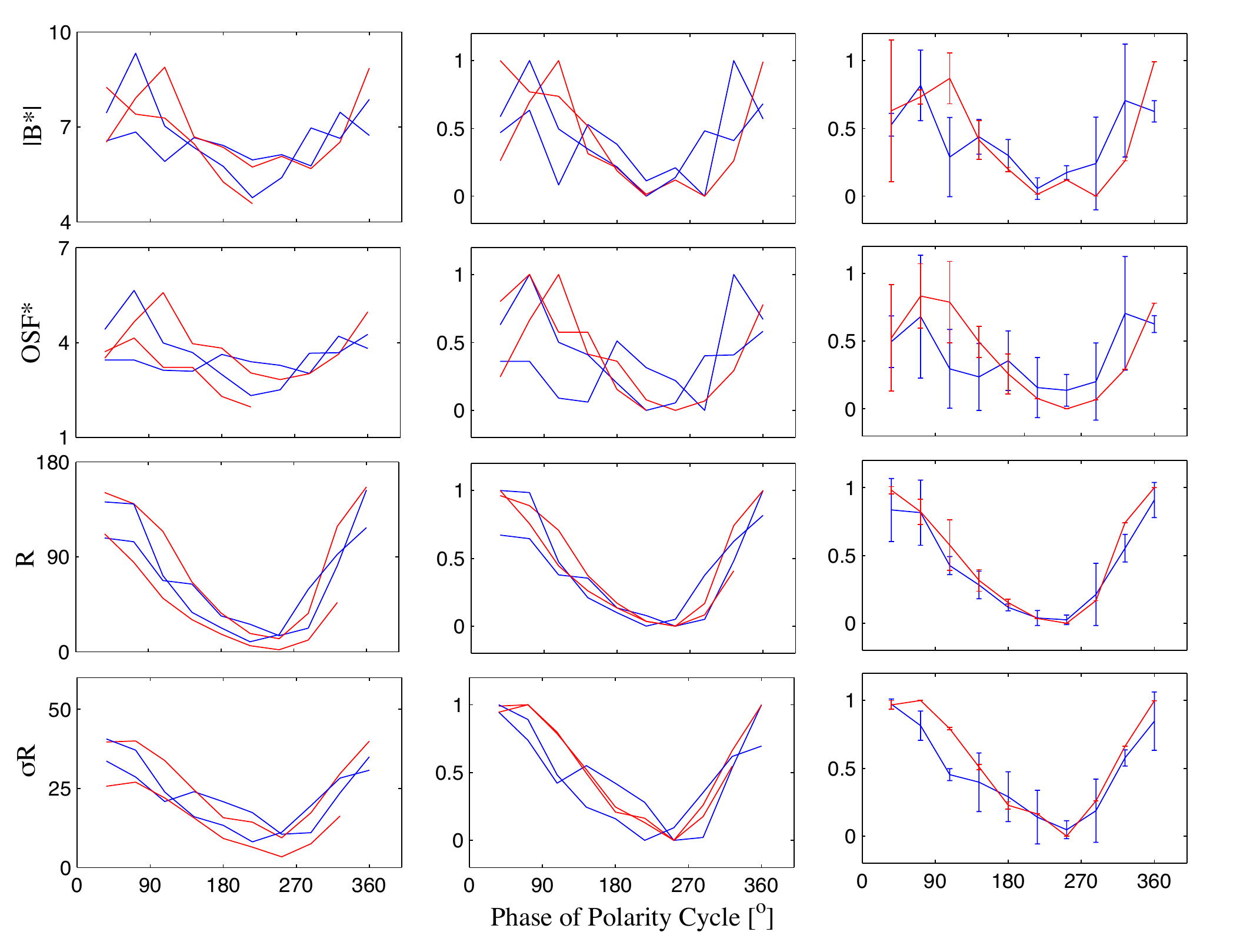}}
\caption{Heliospheric magnetic field magnitude [$|$B$^*$$|$] and open solar flux [OSF$^*$] reconstructed from geomagnetic data, along with monthly sunspot number and monthly standard deviation of daily sunspot number This plot considers the space age period (1965 - 2012). The left column shows the raw data, the middle column shows the data normalised to the maximum and minimum values, and the right column gives the mean and standard deviations of the polarity cycles.}
\label{fig:5}
\end{figure}

Figure 5 shows the space age data only, whereas figure 6 gives the pre-space age data. As can be seen from figure 5, the difference between the qA$<$0 and qA$>$0 cycles during the space age is still present in $|$B$^*$$|$ and OSF$^*$. Due to the reconstructed data having a yearly resolution, the data is binned more coarsely, with fewer data points averaged to produce each data point, which may partly explain the differences in $|$B$^*$$|$ and OSF$^*$ being slightly less pronounced than for $|$B$|$ and OSF in Figure 4. The sunspot numbers and standard deviations are also plotted at this yearly resolution, and show the same trends as found previously for the monthly averages. Note that the geomagnetic data does not cover the final 72$^\circ$ of the polarity cycle phase. This is due to the geomagnetic data only being available up until June 2008, while in this study the main focus times being is the declining phase of the solar cycle.

\begin{figure}[h]\centerline{\includegraphics[width=0.8\textwidth,clip=]{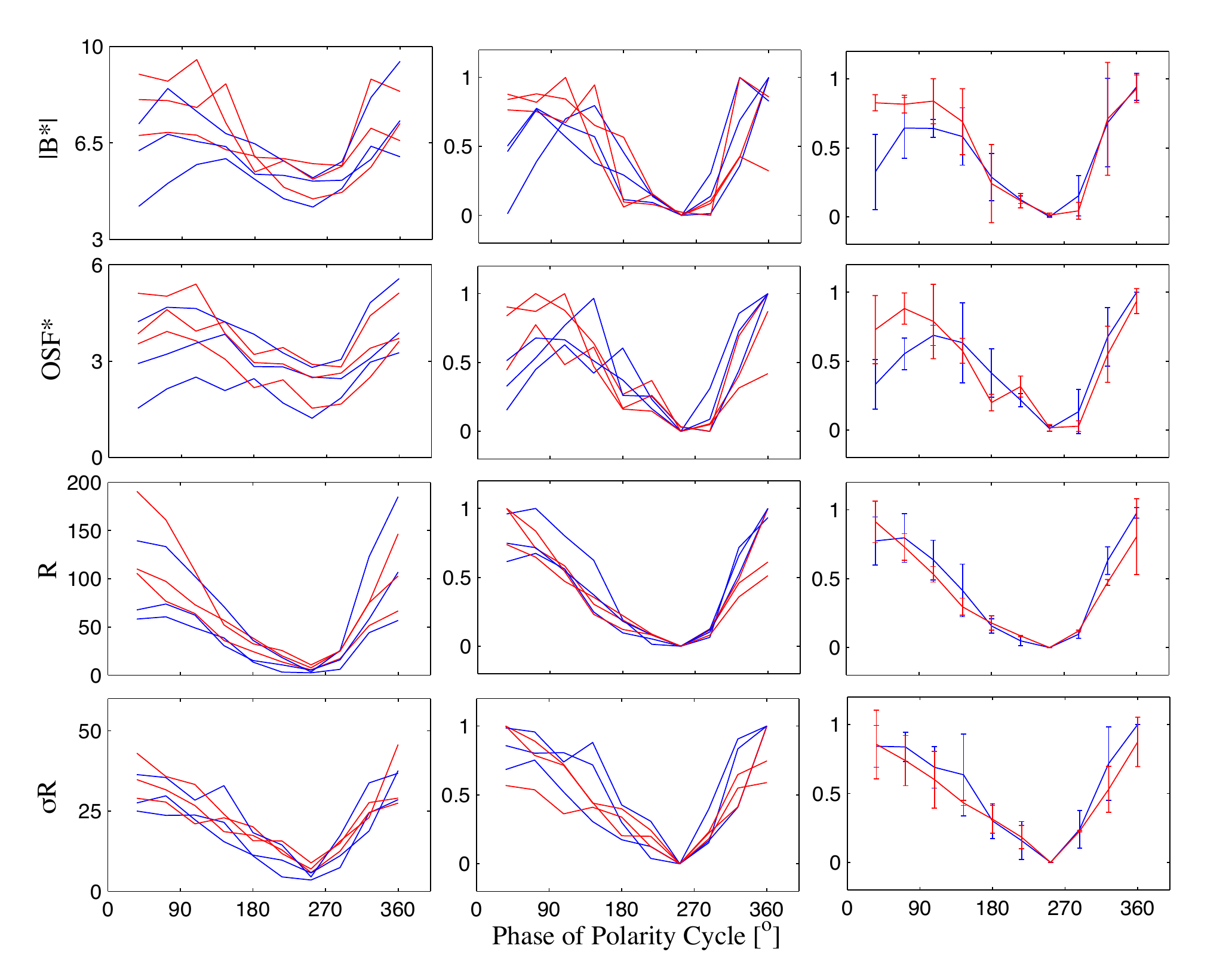}}
\caption{Pre-Space-Age reconstructions of heliospheric magnetic field magnitude [$|$B$^*$$|$] and open solar flux [OSF$^*$]; the sunspot number and standard deviation of daily sunspot number are also shown. The left column shows the raw data, the middle column shows the data normalised and the right is a superposed epoch analysis.}
\label{fig:6}
\end{figure}

Figure 6 shows the same analysis as in Figure 5 but for six pre-space age polarity cycles. The pre-space age $|$B$^*$$|$ and OSF$^*$ do show some slight differences in alternate polarity cycles; however, unlike the space-age, these are earlier in the polarity cycle ({\it i.e.} just after solar maximum) and hence no longer coincident with the Hale Cycle differences in the cosmic ray flux. Furthermore, the standard deviation of daily sunspot numbers with respect to the monthly averages does not show the same variation between alternate polarity cycles in the pre-space age data (in fact qA$>$0 cycles give slightly larger mean values at the relevant $\epsilon$$_p$ instead of the qA$<$0 cycles in Figure 5, but the difference is small compared to the errors).

We now test the sensitivity of the results shown in Figures 5 and 6 to realistic variations on the start and end times of the polarity cycle ({\it i.e.} changes in the time of polarity reversal). Comparison of the polarity reversal times determined from sunspot number and photospheric magnetograph data (Figure 1) suggests a typical uncertainty of around 0.5 years. Thus we perform a Monte Carlo analysis of the difference in heliospheric parameters in qA$<$0 and qA$>$0 cycles to the varying the reversal times by half a year. For each variable shown in Figures 5 and 6, {\it i.e.}, the geomagnetic reconstructions of magnetic field magnitude and OSF, monthly sunspot number, and sunspot variance, we use random numbers to vary the start and end times of each cycle by 0.5 years with a chosen weighting of 50$\%$ chance of no change in start/end time and 50$\%$ of the reversal time changing by 0.5 years (with an equal probability of moving back or forward 0.5 years). For each set of new polarity reversal times, we repeat the same superposed epoch analysis and compute the difference in the qA$>$0 and qA$<$0 parameters. We run this process 1,000 times to get a large spread of cycle start and end times.

\begin{figure}[h]\centerline{\includegraphics[width=0.7\textwidth,clip=]{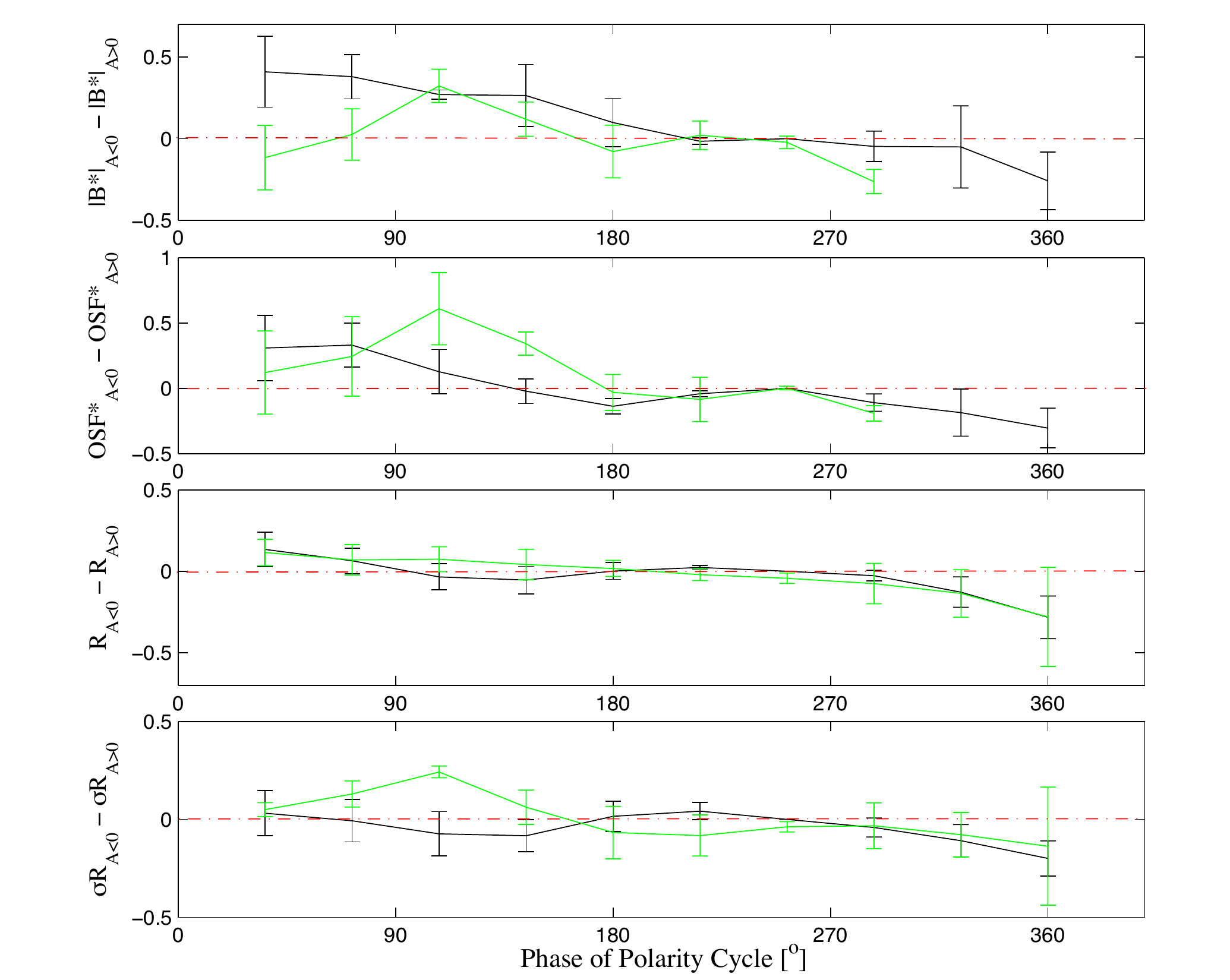}}
\caption{Sensitivity analysis of geomagnetic and sunspot data to varying the start and end dates by plus or minus 0.5 years. Each panel shows the difference between qA$<$0 and qA$>$0 polarity cycles of the following parameters (from top): reconstructed magnetic field intensity, reconstructed open solar flux, sunspot number and sunspot variance. The green line shows the mean of all space-age cycles and the black shows all pre-space age cycles with the error bars represents the standard deviation of all cycles included in the mean.}
\label{fig:7}
\end{figure}

Figure 7 shows the difference in heliospheric parameters between qA$>$0 and qA$<$0 cycles, with space age (pre-space age) in green (black). Shown from top to bottom are geomagnetic reconstructions of the heliospheric magnetic field [$|$B$^*$$|$] and the open solar flux [OSF$^*$], sunspot number [R] and sunspot variance [$\sigma$$_R$], as used in previous plots. Each parameter has been averaged over all available cycles of each polarity and the differences between the means for qA$<$0 and qA$>$0 cycles are plotted. The green lines are the average behaviour of all space age cycles and the black lines are all pre-space cycles. The error bars on each plot are plus or minus the standard deviation of all cycles included in the mean of all samples.

For the heliospheric magnetic field strength [$|$B$^*$$|$] (top panel) both the pre-space age data (black line) and the space-age data (green line) show a difference around $\epsilon$$_p$ = 100$^\circ$ that is larger than the error bars and so can be considered significant. This difference is significant at all phases of the declining phase of the sunspot cycle ($\epsilon$$_p$ $<$ 100$^\circ$) for the pre-space age data but not for shortly after solar maximum in the space-age data. This means that shortly after sunspot maximum, $|$B$^*$$|$ during qA$<$0 cycles is larger than for qA$>$0 cycles in the pre-space age data, which is the opposite of the space-age data. The result that there is no significant difference between the two polarity cycles for sunspot number [R] is found not to be sensitive to the start and end times of cycles used.

On the other hand, for both the OSF$*$ and sunspot number variability [$\sigma$$_R$], the result that the declining phases of space age cycles show an increase in these parameters during qA$<$0 over qA$>$0 but do not show a difference in pre-space age data, holds when subjected to sensitivity testing. This is shown as the peak in the space age data above the pre-space age is more than zero by more than the error bar.

Another result to note is the difference at the start of the polarity cycle seen in the geomagnetic $|$B$^*$$|$ data. Here we see $|$B$^*$$|$ during qA$<$0 dominating $|$B$^*$$|$ during qA$>$0 during pre-space cycles but the opposite being true for space age cycles. This result may warrant further work but is not seen in any other variable.


\section{Discussion and Conclusions}

As cosmic rays are modulated by the heliospheric magnetic field and heliospheric current sheet tilt, the 11-year cycle is also seen in cosmic ray records. In addition to the solar cycle, cosmic ray time series display a strong 22-year Hale Cycle, which has been attributed to differing drift patterns and diffusion (particularly at solar maximum) during positive and negative solar field polarities ({\it e.g.} \opencite{Jokipii77}; \opencite{Ferreira04}). It has been argued that this results in the rise to the cosmic ray peak earlier during qA$>$0 cycles than for qA$<$0 cycles, and gives the time series a “flat-topped” and “peaked” appearance respectively. However, an increasing number of results are not consistent with this concept. For example, \inlinecite{Richardson99} and \inlinecite{Gil08} have found that recurrent decreases in cosmic ray fluxes were considerably larger when qA$>$0, whereas the drift theory suggests that they should be larger for qA$<$0, when cosmic rays should be entering by drifting inward along the HCS. In addition other studies have found differences between qA$>$0 and qA$<$0 in heliospheric current sheet tilt \cite{Cliver01} and in open solar flux \cite{Rouillard04} which offer alternative explanations of the 22-year cycle in cosmic ray fluxes.

In this study, we have separated space-age solar and heliospheric data into “polarity cycles”, defined as intervals between polar solar polarity reversals, thus approximately spanning solar maximum to solar maximum.

The results show a significant difference in the behaviour of HCS inclination, heliospheric magnetic field magnitude, and open solar flux between qA$>$0 cycles and qA$<$0 cycles. This difference is only significant during the first half of the polarity cycle, which corresponds to the declining phase of the solar cycle, the period responsible for a large part of the difference in GCR fluxes in qA$>$0 and qA$<$0 cycles. The standard deviation in daily sunspot number also gives a significant difference between qA$>$0 and qA$<$0 cycles during the same period, suggesting a greater prevalence of active longitudes during this phase of qA$<$0 cycles. This is in agreement with the increased HCS inclination throughout this period. The presence of more active longitudes giving greater HCS inclination means that there will be regular fast/slow stream interfaces extending over large latitudinal ranges, which was shown to be an effective way of shielding cosmic rays in a case study by \inlinecite{Rouillard07}.

We suggest that the 22-year cycle in GCR flux may be partly due to direct heliospheric modulation, although drift effects (\opencite{Jokipii77}; \opencite{Ferreira04}) will still play a role, particularly during the end of the polarity cycle ({\it i.e.}, the rise phase of the solar cycle), when differences in heliospheric parameters are less apparent. Of course, while changes in heliospheric structure are coincident with the differing behaviour in cosmic ray  flux in alternate polarity cycles, it still remains to be shown that they are of sufficient magnitude to effect the required modulation. To do this will, however, require a significant modelling effort.

The above conclusions relate to the space-age era for which there are {\it in-situ} observations of interplanetary parameters and magnetograph data from which the start/end times of the polarity cycle and the HCS tilt index can be derived. As this data has been taken during a grand solar maximum \cite{Lockwood09} it is not necessarily the case that the conclusions will have been true in the less active times prior to the grand maximum. To test this, the open solar flux and near-Earth magnetic field reconstructed from geomagnetic activity data have been used, employing the sunspot number variations to define the start/end times of the polarity cycle. The data was divided into the space-age era (1965 and after corresponding to the first study) and pre- space age data before 1965.

The reconstructed data from the space-age era support the above findings of the study using direct observations, but it is also noted that the differences between polarity cycles are considerably smaller before the space age. Using geomagnetic reconstructions of heliospheric magnetic field magnitude and open solar flux, it was shown that for the period of 1905 - 1965, the opposite polarities do not give such differing patterns during the declining phase of the solar cycle. In particular, the variability is sunspot numbers is greatly reduced. One source of uncertainty that we address is that before the space age we have to use the sunspot number variation to define the start/end time of the polarity cycles. A sensitivity study that adds the uncertainty in these inferred times shows that the result is robust for the open solar flux [OSF] and the variability of the sunspot number [$\sigma$$_R$]. However, this uncertainty means that we cannot be certain that the polarity effect on the near-Earth heliospheric field strength (at the phase of the polarity cycle when the polarity effect on GCR is greatest in the space-age era) is different in the pre-space age data, compared to the space-age era.

The data suggests that the polarity-dependent effect on cosmic rays before the recent grand solar maximum was most likely to be restricted to the drift effects and not as marked as they have been in recent data. This is consistent with cosmogenic isotope data, which, in general, do not show strong 22-year Hale cycle variations \cite{Usoskin08}.



 \begin{acks}
 
 We are grateful to the Space Physics Data Facility (SPDF) of NASA’s Goddard Space Flight Centre for combining the data into the OMNI 2 data set, which was obtained via the GSFC/SPDF OMNIWeb interface at {http://omniweb.gsfc.nasa.gov} and to the Marshall Space Flight Centre for the Sunspot Number data obtained from MSFC at {http://solarscience.msfc.nasa.gov/greenwch.shtml}. We also thank the Bartol Research Institute of the University of Delaware for the neutron monitor data from McMurdo, which is supported by NSF grant ATM-0527878 and J. T. Hoeksema of Stanford University for WSO magnetograms. The work of SRT is supported by a studentship from the UK’s Natural Environment Research Council (NERC).
 
 \end{acks}
 



 \bibliographystyle{spr-mp-sola}
 \bibliography{Hale}  




\end{article} 
\end{document}